\documentstyle[preprint,aps,epsf]{revtex}

\begin{document}

\preprint{\font\fortssbx=cmssbx10 scaled \magstep2
\hfill$\vcenter{\hbox{\bf FTUV/96-82}
		\hbox{\bf IFIC/96-91}
                \hbox{\bf IFUSP-P 1251}
                \hbox{\bf hep-ph/9612254}
                \hbox{\bf December 1996}
}$}

\title{Z Physics Constraints on Vector Leptoquarks}

\author{O.\ J.\ P.\ \'Eboli
\footnote{ Permanent address: Instituto de F\'{\i}sica da
  USP, C.\ P.\ 66.318, 05315-970 S\~ao Paulo, SP, Brazil. E-mail:
  eboli@fma.if.usp.br}, and M.\ C.\ Gonzalez-Garcia
\footnote{E-mail: concha@evalvx.ific.uv.es}
}
\address{ Instituto de F\'{\i}sica Corpuscular - IFIC/CSIC,
Departament de F\'{\i}sica Te\`orica \\
Universitat de Val\`encia, 46100 Burjassot, Val\`encia, Spain
}
\author{J.\ K.\ Mizukoshi
\footnote{ E-mail: mizuka@fma.if.usp.br} }

\address{Instituto de F\'{\i}sica,  Universidade de S\~ao Paulo, \\
Caixa Postal 66.318, 05315-970 S\~ao Paulo, Brazil.}

\maketitle
\thispagestyle{empty}

\begin{abstract}
 \baselineskip 0.5 cm
  We analyze the constraints on vector leptoquarks coming from
  radiative corrections to $Z$ physics.  We perform a global fitting
  to the LEP data including the oblique and non-universal
  contributions of the most general effective Lagrangian for vector
  leptoquarks, which exhibits the $SU(2)_L \times U(1)_Y$ gauge
  invariance. We show that the $Z$ physics leads to stronger bounds on
  second and third generation vectors leptoquarks than the ones
  obtained from low energy and the current collider experiments.
\end{abstract}

\begin{center}
Submitted to Physics Letters B
\end{center}
\newpage

\section{Introduction}

A large number of extensions of the standard model (SM) predicts the
existence of color triplet particles with spin $0$ or $1$, which carry
simultaneously leptonic and baryonic number; the so-called
leptoquarks. Leptoquarks are present in models that treat quarks and
leptons on the same footing, and consequently allow quark-lepton
transitions. This class of models includes composite models
\cite{comp}, grand unified theories \cite{gut}, technicolour
\cite{tech}, and superstring-inspired \cite{rizzo} models.

The impressive success of the SM  \cite{sm} poses strong limitations on the 
possible
forms of new physics even when this one cannot be produced directly.
In the present work we make use of this fact to obtain bounds on vector 
leptoquarks through
their contributions to the radiative corrections to the $Z$ physics.
We evaluate the one-loop contribution due to leptoquarks to all LEP
observables, keeping only its non-analytical part.  Then, the limits on
the leptoquark masses and couplings are obtained through a global
fitting to all available LEP data. Contrary to what happens to scalar
leptoquarks \cite{nois}, $Z$ physics sets bounds on first and
second generation leptoquarks, in addition to the third generation
ones. The bounds on leptoquarks coupling to the first generation are
less restrictive than the ones derived from low energy experiments,
except for very small couplings of the leptoquarks to the
fermions. Notwithstanding, we obtain the strongest available limits on
vector leptoquarks that couple to the second and third families.

Since the discovery of a leptoquark is an undeniable signal
of new physics, there have been also many direct searches for leptoquarks
in accelerators.  The four LEP experiments, assuming that the $Z$ can decay
into a pair of on-shell leptoquarks, established a lower bound $
M_{\text{lq}} \gtrsim 44$ GeV for scalar leptoquarks \cite{lep}, and a
similar bound should hold for vector ones.  The HERA collaborations
\cite{hera} obtained limits on the masses and couplings of first
generation leptoquarks for masses up to 275 GeV, depending on the
leptoquark type and couplings. The two Tevatron collider experiments,
CDF and D0, searched for leptoquark pairs leading to charged dileptons plus
one dijet. For leptoquarks decaying exclusively into charged
lepton and a jet, these collaborations constrained the scalar
leptoquark masses to be larger than 180 (94) GeV for second and third
generation leptoquarks, while similar limits exist for vector
leptoquarks depending on their anomalous chromomagnetic moments
\cite{pp}. There have also been some studies of the possibility of 
observing vector leptoquarks in the future $pp$ \cite{fut:pp} and
$e^+e^-$ \cite{fut:ee} colliders.

Low energy experiments give rise to strong constraints on
leptoquarks, unless their interactions are carefully chosen
\cite{shanker,fcnc}. In order to evade the bounds from proton decay,
leptoquarks are required not to couple to diquarks. Moreover, 
to avoid the appearance of leptoquark induced FCNC, leptoquarks are
assumed to couple only to a single quark family and only one lepton
generation. Nevertheless, there still exist low-energy limits on
leptoquarks. Helicity suppressed meson decays restrict the couplings
of leptoquarks to fermions to be chiral. Moreover, residual FCNC,
atomic parity violation, and meson decay \cite{leurer,apv} constrain
the first generation leptoquarks to be heavier than
$0.5$--$1.5$ TeV when the coupling constants are equal to the
electromagnetic coupling $e$.


\section{Effective Interaction and Analytical Expressions}
\label{l:eff}

Our starting point is the most general effective Lagrangian for vector leptoquarks 
invariant under the gauge symmetry $SU(2)_L \times U(1)_Y$,  
and baryon ($B$) and lepton ($L$) number conserving  \cite{buch}. This last condition is
needed to comply with the strong
bounds coming from the proton lifetime experiments;
\begin{equation}
{\cal L} = {\cal L}^f_{|F|=2}+{\cal L}^f_{|F|=0}+
{\cal L}^{\gamma, Z, W} + {\cal L}_{\text{dipole}} \; ,
\label{lalepto}
\end{equation}
where the interactions with the fermions are described by
\begin{eqnarray}
{\cal L}^f_{|F|=2} &=&
g_{2L}~ (V^L_{2\mu})^T~ \bar{d}^c_R{\gamma}^{\mu}i{\tau}_2l_L~ +~
g_{2R}~ \bar{q}^c_L{\gamma}^{\mu}i{\tau}_2e_R~ V^R_{2\mu}
\nonumber \\
&& +~ \tilde{g}_{2L}~ (\tilde{V}^L_{2\mu})^T~
\bar{u}^c_R{\gamma}^{\mu}i{\tau}_2l_L~ +~ \mbox{h.c.} \; ,
\label{f2}
\\
{\cal L}^f_{|F|=0} &=& 
h_{1L}~ \bar{q}_L{\gamma}^{\mu}l_L~  U_{1\mu}^L~
+~ h_{1R}~ \bar{d}_R{\gamma}^{\mu}e_R~ U_{1\mu}^R~
+\tilde{h}_{1R}~ \bar{u}_R{\gamma}^{\mu}e_R~ \tilde{U}_{1\mu}^R
\nonumber \\
&&+~ h_{3L}~ \bar{q}_L\vec{\tau}{\gamma}^{\mu}l_L~ \vec{U}_{3\mu}^L
~+~ \mbox{h.c.} \; .
\label{f0}
\end{eqnarray}
Here $F=3B+L$, $q$ ($\ell$) stands for the left-handed quark (lepton)
doublet, and $u_R$, $d_R$, and $e_R$ are the singlet components of the
fermions. We denote the charge conjugated fermion fields by
$\psi^c=C\bar\psi^T$ and we omitted in Eqs.\ (\ref{f2}) and (\ref{f0})
the flavour indices of the couplings to fermions and leptoquarks. The
leptoquarks $U^{L(R)}_1$ and $\tilde{U}^R_1$ are singlets under
$SU(2)_L$, while $V^{L(R)}_2$ and $\tilde{V}^L_2$ are doublets, and
$U_3$ is a triplet.  Furthermore, we assumed in this work that the
leptoquarks belonging to a given $SU(2)_L$ multiplet are degenerate in
mass, with their mass denoted by $M_{\text{lq}}$.  For the sake of
simplicity, we also assumed that the leptoquarks couple to leptons and
quarks of the same family.

It is convenient to summarize each term in the interaction
lagrangians (\ref{f2}) and (\ref{f0}) as 
\begin{equation}
g_{\text{lq},X}~ \sum_{j, q, \ell}~ M^j_{q \ell}~ \bar{q^{(c)}} 
\gamma^\mu P_X \ell
~V^j_\mu + \text{h.c.} \; ,
\end{equation}
where the sum is over all quarks, leptons, leptoquarks in the
multiplets.  $P_X$, with $X = L$ or $R$, stands for the helicity
projectors, while the matrices $M^j_{q\ell}$ can be easily obtained
by comparing the above expression with Eqs.\ (\ref{f2}) and (\ref{f0}).

Local invariance under $SU(2)_L \times U(1)_Y$ implies that
leptoquarks also couple to the electroweak gauge bosons. To obtain the
couplings to $W^\pm$, $Z$, and $\gamma$, we substituted $\partial_\mu$
by the electroweak covariant derivative in the leptoquark kinetic
Lagrangian, resulting in ${\cal L}^{\gamma, Z, W}$.
\begin{eqnarray}
{\cal L}^{\gamma, Z, W} &=&
\sum_{\mbox{lq}} \left [ -\frac{1}{2}~ 
G_{\mu \nu}^{\dagger}G^{\mu \nu}+
M_{\text{lq}}^2~ {\Phi}^{\mu \dagger}{\Phi}_{\mu} \right ] \;  ,
\end{eqnarray}
where ${\Phi}_{\mu}$ stand for any vector leptoquark multiplet.  The
leptoquark field strength tensor is $G_{\mu \nu} =
D_{\mu}{\Phi}_{\nu}-D_{\nu}{\Phi}_{\mu}$, with the covariant
derivative
\[
D_{\mu} =
{\partial}_{\mu}-\frac{ie}{\sqrt{2}s_W}[W^+_{\mu}I^++W^-_{\mu}I^-]-ieQ_Z
Z_{\mu}+ieQ_{\gamma}A_{\mu},
\]
where $Q_\gamma$ is the leptoquark electric charge in units of the
positron charge, $Q_Z = (I_3-Q_{\gamma}s_W^2)/s_Wc_W$, $I^{\pm} = I_1
\pm iI_2$, and the $I_a$'s are the generator of $SU(2)_L$ for the
leptoquark multiplet. $s_W$ ($c_W$) stands for the sine (cosine) of
the weak mixing angle.

It is possible to write a further contribution to the vector
leptoquark lagrangian that is also invariant under $SU(2)_L
\otimes U(1)_Y$
\begin{equation}
{\cal L}_{\text{dipole}} 
= -i \left [ \frac{\kappa}{2}~ \frac{e}{s_W}~ W_{\mu \nu}^{a}~
(I_a)_{bc}~
-~ \frac{{\kappa}'}{2}~ \frac{e}{c_W}~ 
\biggl(\frac{Y}{2}\biggr)_{b c}~ B_{\mu \nu} \right ]~
\left ({\Phi}^{\dagger b \mu}{\Phi}^{c \nu}-{\Phi}^{\dagger b \nu}
{\Phi}^{c \mu} \right ) \; .
\label{laadicional}
\end{equation}

In this work we employed the on-shell-renormalization scheme, adopting
the conventions of Ref.\ \cite{hollik}. We used as inputs the fermion
masses, $G_F$, $\alpha_{\text{em}}$, and the $Z$ mass, hence, the
electroweak mixing angle is a derived quantity defined through $\sin^2
\theta_W = s_W^2 \equiv 1 - M^2_W / M^2_Z$.  Close to the $Z$
resonance, the physics can be summarized by the effective neutral
current
\begin{equation}
J^Z_\mu =  \left ( \sqrt{2} G_\mu M_Z^2 \rho_f
\right )^{1/2} \left [ \left ( I_3^f - 2 Q^f s_W^2 \kappa_f \right )
\gamma_\mu - I_3^f \gamma_\mu \gamma_5 \right ] \; ,
\label{form:nc}
\end{equation}
where $Q^f$ ($I_3^f$) is the fermion electric charge (third component
of weak isospin).  The form factors $\rho_f$ and $\kappa_f$ have universal
contributions, {\sl i.e.} independent of the fermion species, as well
as non-universal parts:
\begin{equation}
\rho_f  = 1 + \Delta \rho_{\text{univ}} +
\Delta \rho_{\text{non}} \; , \;\;\;\;\;\;\;\;\;\;\;\;\;
\kappa_f = 1 + \Delta \kappa_{\text{univ}} +
\Delta \kappa_{\text{non}} \; .
\end{equation}

\begin{figure}
\hskip 1cm
\parbox[c]{1.in}{\epsfxsize=6cm \epsfbox{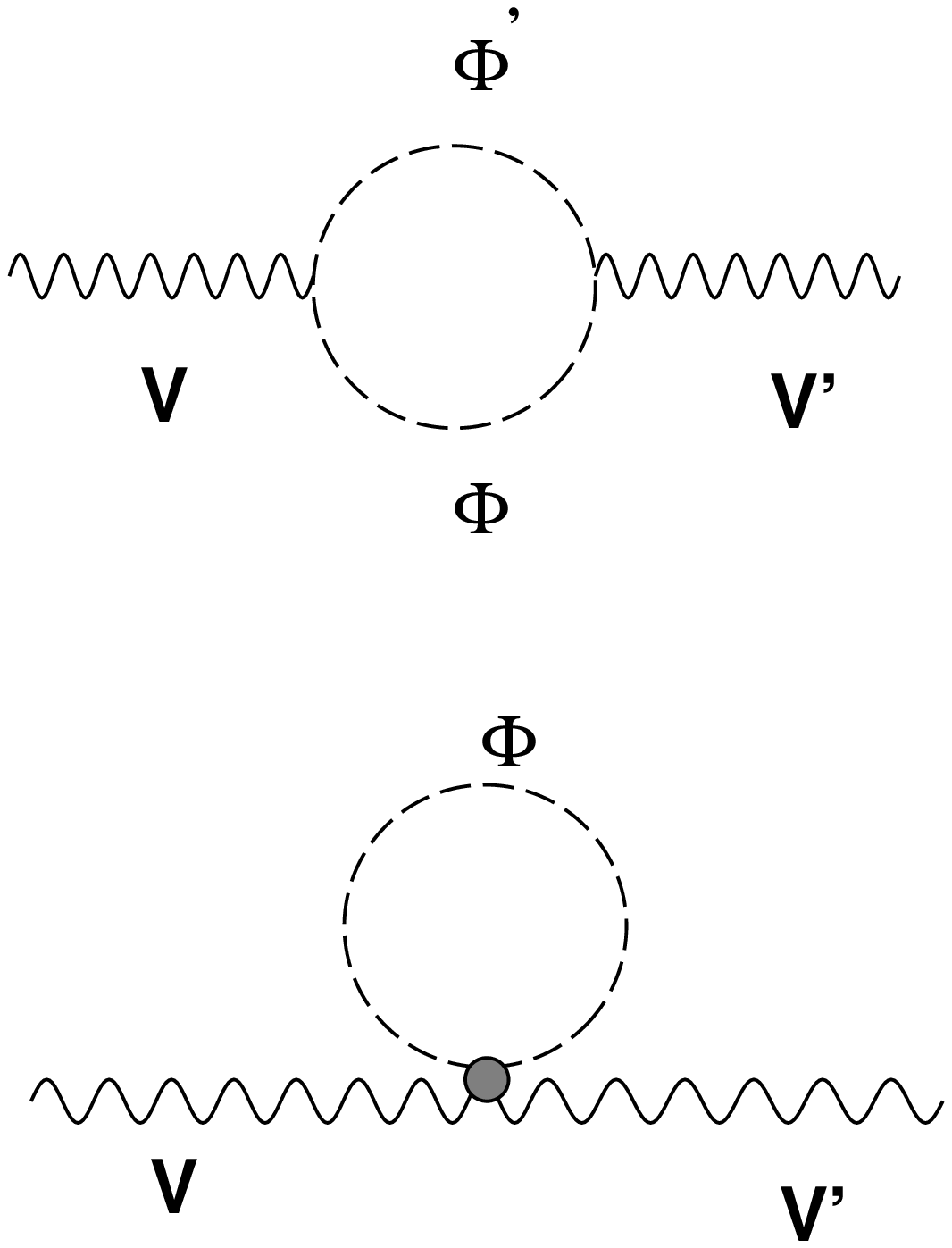} }
\hskip 1cm
\parbox[c]{1.7in}{\epsfysize=9cm \epsfbox{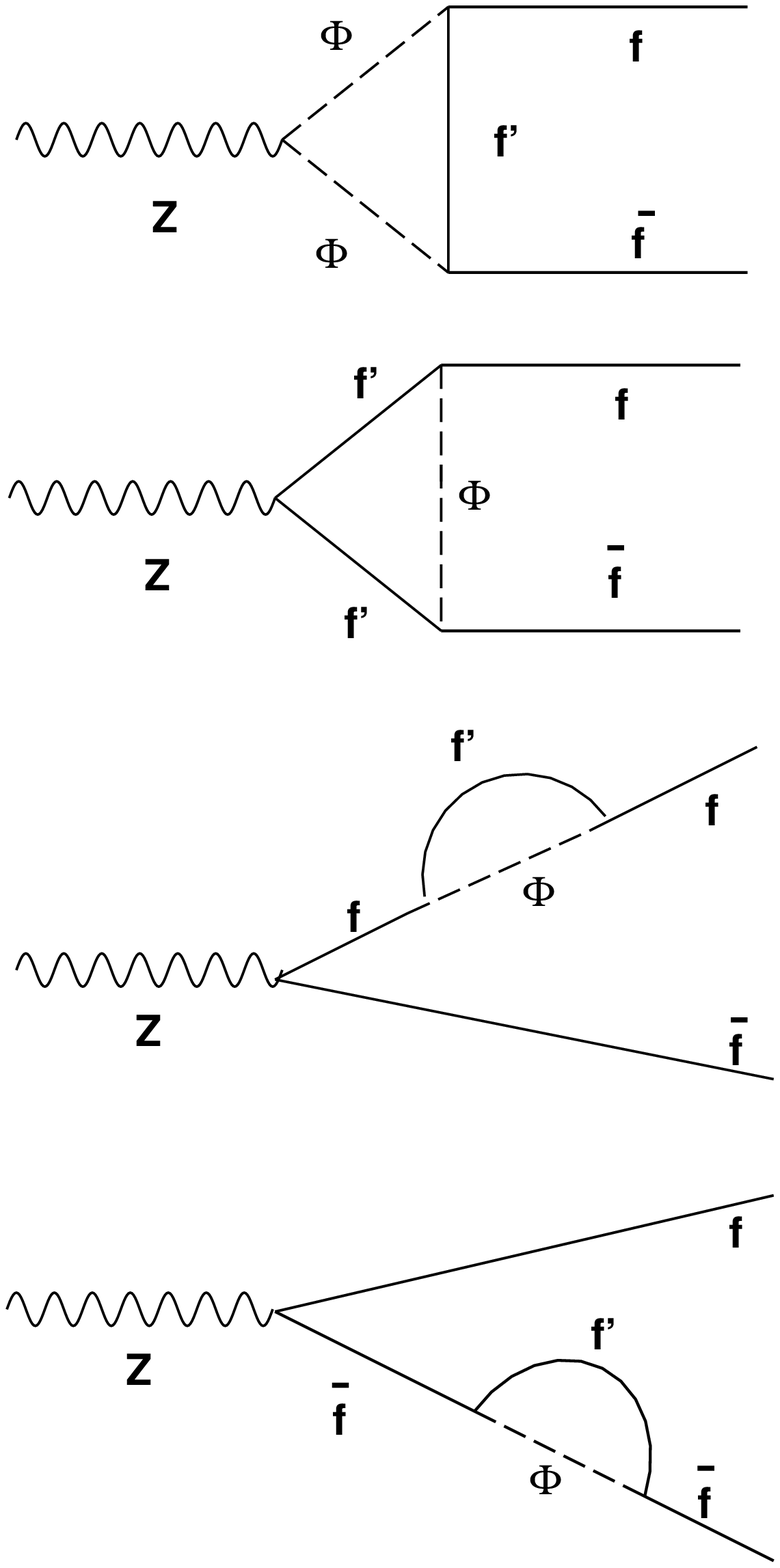}
}
\caption{Feynman diagrams leading to leptoquark contribution to
vector-boson self-energies (left column) and to the vertex
$Z$--$f$--$\bar{f}$ (right column).}
\label{prop}
\end{figure}

Leptoquarks modify the physics at the $Z$ pole through their
contributions to both universal \cite{obli} and non-universal
corrections. The universal contributions, displayed in Fig.\
\ref{prop}, can be expressed in terms of the unrenormalized vector
boson self-energy ($\Sigma$) as
\begin{eqnarray}
\Delta \rho^{\text{lq}}_{\text{univ}}(s) &=&
-\frac{\Sigma^Z_{\text{lq}}(s)-\Sigma^Z_{\text{lq}}(M_Z^2)}{s-M_Z^2}
+\frac{\Sigma^Z_{\text{lq}}(M_Z^2)}{M_Z^2}
-\frac{\Sigma^W_{\text{lq}}(0)}{M_W^2} - 2 \frac{s_W}{c_W}~
\frac{\Sigma^{\gamma Z}_{\text{lq}}(0)}
{M_Z^2}
\; ,\\
\Delta \kappa^{\text{lq}}_{\text{univ}} &=& - \frac{c_W}{s_W}~
\frac{\Sigma^{\gamma Z}_{\text{lq}}(M_Z^2)}{M_Z^2}
- \frac{c_W}{s_W}~
\frac{\Sigma^{\gamma Z}_{\text{lq}}(0)}{M_Z^2}
+\frac{c_W^2}{s_W^2} \left[ \frac{\Sigma_{\text{lq}}^Z(M_Z^2)}{M_Z^2}-
\frac{\Sigma_{\text{lq}}^W(M_W^2)}{M_W^2}\right]
\; ,
\end{eqnarray}
where $s$ is the square of the vector-boson four momentum.

Corrections to the vertex $Z f \bar{f}$ give rise to non-universal
contributions to $\rho_f$ and $\kappa_f$.  Leptoquarks affect these
couplings of the $Z$ through the diagrams also given in Fig.\
\ref{prop} whose results we parametrize as
\begin{equation}
i \frac{e}{2 s_W c_W} \left [ \gamma_\mu F_{V\text{lq}}^{Zf} - \gamma_\mu \gamma_5
F_{A\text{lq}}^{Zf} + I_3^f \gamma_\mu (1 - \gamma_5) \frac{c_W}{s_W} ~
\frac{\Sigma^{\gamma Z}_{\text{lq}}(0)}{M_Z^2} \right ] \; .
\end{equation}
This leads to
\begin{equation}
\Delta \rho^{\text{lq}}_{\text{non}} = 2~\frac{F_{A\text{lq}}^{Zf}}{a_f}(M_Z^2)
\;,  \;\;\;\;\;\; 		 
\Delta \kappa^{\text{lq}}_{\text{non}} = - \frac{1}{2 s_W^2  Q^f} \left [
F_{V\text{lq}}^{Zf}(M_Z^2) - \frac{v_f}{a_f}~ F_{A\text{lq}}^{Zf}(M_Z^2)
\right ]
\; ,
\end{equation}
where $a_f = I_3^f$ and $v_f = I^f_3 - 2 Q^f s_W^2$.

In order to evaluate the relevant Feynman diagrams we used dimensional
regularization \cite{reg:dim} and took the external fermion masses to
be zero. Then, we retained only the leading non-analytical contributions
from the loop diagrams -- that is, the contributions that are relevant
for our analysis were obtained by the substitution
\begin{equation}
\frac{2}{4-d} \rightarrow {\rm{log}}\;\frac{\Lambda^2}{M_Z^2}\; ,
\nonumber
\end{equation}
where $\Lambda$ is the energy scale which characterizes the appearance
of new physics, and we dropped all other terms.2

The contribution of the vector leptoquarks to the vector-boson
self-energies depends exclusively on their gauge couplings and it is
given by
\begin{equation}
\begin{array}{ll}
\Sigma^W_{\text{lq}}(q^2) = \frac{\displaystyle 1}{\displaystyle s_W^2}
~{\displaystyle \sum_j}
\left ( I_3^j \right )^2~ q^2~ \mbox{I}(q^2)
\;,  \;\;\;\;\;\;\;\;\;\;\;& 
\Sigma^\gamma_{\text{lq}}(q^2) = {\displaystyle \sum_j} \left ( Q_\gamma^j \right )^2~
q^2~ \mbox{I}(q^2)
\; , 
\\
\Sigma^Z_{\text{lq}}(q^2) = {\displaystyle \sum_j} \left ( Q_Z^j \right )^2~
q^2~ \mbox{I}(q^2)
\; , &
\Sigma^{\gamma Z}_{\text{lq}}(q^2) = -{\displaystyle \sum_j}  Q_\gamma^j Q_Z^j~ q^2~
\mbox{I}(q^2)
\; ,
\end{array}
\end{equation}
where the sum is over all members of a given leptoquark multiplet and 
we defined
\begin{equation}
\mbox{I}(q^2) \equiv \frac{\alpha_{\text{em}}}{48 \pi}~ N_c~ 
\ln \left ( \frac{\Lambda^2}{M_{\text{lq}}^2} \right )~ 
\left [ -12 -72 \kappa + \frac{q^2}{M_{\text{lq}}^2} \left (
4 + 12 \kappa - 2 \kappa^2 \right ) + \kappa^2 \frac{q^4}{M_{\text{lq}}^4}
\right ] \; ,
\end{equation}
with $N_c=3$ being the number of colors. For simplicity we assumed
that $\kappa = \kappa^\prime$.

The effect of the vector leptoquarks to the couplings of the $Z$ to a
lepton pair $\ell\bar{\ell}$ is
\begin{equation}
F_{V, \text{lq}}^\ell = \mp F_{A, \text{lq}}^\ell
 = \frac{g_{\text{lq},X}^2}{192 \pi^2} s_W c_W N_c~
\sum_{j,q} {M_{\ell q}^j}^\dagger M_{q \ell}^j ~Q^j_Z~ 
 q^2~ \left [ \kappa \left ( 3 \frac{m_q^2}{M_{\text{lq}}^2}
- \frac{q^2}{M_{\text{lq}}^2} -6 \right ) - 10 \right ] \;
\ln \left ( \frac{\Lambda^2}{M_{\text{lq}}^2} \right ) 
\; ,
\label{z:ll}
\end{equation}
for both $|F|=2$ and $F=0$ vector leptoquarks.  
$N_c = 3$, $m_q$ is the mass of the quark running in the loop,
and the $+$ ($-$) corresponds to left- (right-) handed leptoquarks.
For $F=0$ leptoquarks, in order to obtain the $Zq\bar{q}$ vertex correction 
we just have to
change $\ell \Longleftrightarrow q$ in the above expression and set
$N_c=1$. For $|F|=2$ leptoquarks one also has to change 
$g_{\text{lq},X}\Longleftrightarrow g_{\text{lq},-X}$ where we denote $R=-L$.
 
In order to check the consistency of our calculations, we verified
that the effect of leptoquarks to the $\gamma f \bar{f}$ vertex
vanishes at zero momentum, which is used as one of the renormalization
conditions in the on-shell scheme. This vertex function can be
obtained from (\ref{z:ll}) by the substitutions $Q_Z^j \Rightarrow
-Q^j_\gamma$, $e/2s_Wc_W \Rightarrow -e$. As we can see from Eq.\
(\ref{z:ll}), the leading leptoquark contribution to
the vertex function $\gamma f \bar{f}$ is proportional to $q^2$, and
hence, the QED Ward identities \cite{ward} are satisfied, since the
fermion electric charges at $q=0$ are left unchanged.

Leptoquark corrections for the vertex $W\,$$\ell\,$$\nu_\ell$ ($\ell =
e$ or $\mu$) at low energy contribute to $\Delta r$, see for instance
the first reference in \cite{nois}, and consequently to $\Delta
\rho_{\text{univ}}$. However, the leading non-analytical
correction is proportional to the
 mass of the quark running in the
loop and  therefore it vanishes since we assume that the
leptoquark couples to the same generation of leptons and quarks.


\section{Results and Discussion}

We performed a global fit to all LEP data including both universal and
non-universal contributions. In Table \ref{LEPdata} we show the most
recent combined results of the four LEP experiments. These
results are not statistically independent and the correlation matrix
can be found in \cite{sm}. We expressed the theoretical predictions to
these observables in terms of $\kappa^f$, $\rho^f$, and $\Delta r$,
with the SM contributions being obtained from the program ZFITTER
\cite{zfit}. In order to perform the global fit we constructed the
$\chi^2$ function and minimized it using the package MINUIT. In our
fit we used five parameters, three from the SM ($m_{\text{top}}$,
$M_H$, and $\alpha_s(M_Z^2)$) and two new ones: $M_{\text{lq}}$ and
$g_{\text{lq}}$. We constrained the value of $m_{\text{top}}$ to 
lie in the range allowed by the Tevatron experiments \cite{mtop}.

\begin{figure}
\epsfysize=8.5cm 
\begin{center}
\leavevmode \epsfbox{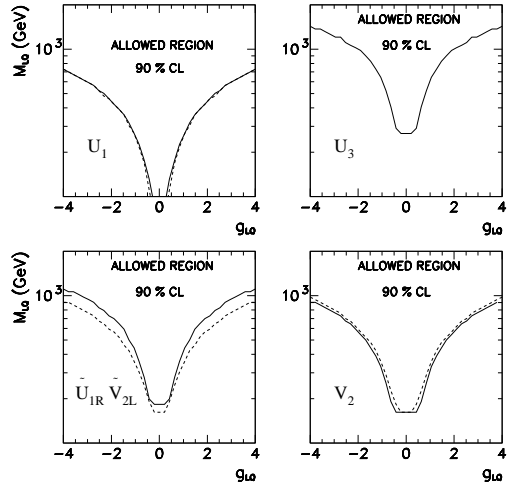}
\end{center}
\caption{Allowed regions (90 \% CL) in the plane 
$M_{\text{lq}}$--$g_{\text{lq}}$
for third generation vector leptoquarks, $\Lambda=2$ TeV, and $\kappa =
\kappa^\prime =0$.  The values of all other parameters were allowed to vary. 
The solid (dashed) lines stand for left(right)-handed leptoquarks.}
\label{contour:1}
\end{figure}

We present in Fig.\ \ref{contour:1} the 90 \% CL allowed regions in the
plane $M_{\text{lq}}$--$g_{\text{lq}}$ for third generation
leptoquarks, assuming $\Lambda = 2$ TeV and $\kappa=\kappa^\prime =
0$. Similar results hold for leptoquarks coupling to the other
generations.  As we can see from this figure, the bounds become more
stringent as the leptoquark isospin grows. This happens because the
leptoquark contribution to the form factors $F_{V(A)}^f$ increases with
the leptoquark isospin. It is interesting to notice that the bounds on
right-handed and left-handed leptoquarks are basically the same since
they lead to similar contributions to $F_{V(A)}^f$.

Low energy bounds on vector leptoquarks can be evaded provided the
couplings leptoquark--quark--lepton are sufficiently weak, because
they rely on the four fermion effective interactions generated by
leptoquarks. On the other hand, the $Z$ physics can lead to bounds on
leptoquarks, even if the coupling $g_{\text{lq}}$ is very small, due
to their oblique contributions. In Table \ref{tab:1} we display the
generation independent bounds (95 \% CL) that can be obtained assuming
$g_{\text{lq}}=0$, which allow us to access the importance of their
universal contributions.  These bounds are weaker than the present
experimental limits for first generation leptoquarks, however, they
are stronger the available experimental limits for second and third
generation leptoquarks.

We show in Table \ref{tab:2} the 95 \% CL lower limits on the masses of
second and third generation leptoquarks for several values of the
couplings $g_{\text{lq}}$ and $\kappa=\kappa^\prime$. Since the limits
for leptoquarks coupling to the first generation are weaker than the
ones drawn for the low energy experiments, we don't exhibit them.  In
order to generate this table, we fixed $m_{\text{top}}= 175$ GeV and
$\Lambda = 2$ TeV, and we varied $\alpha_S(M_Z)$ in the interval
$0.121\pm0.005$ and $M_H$ in the range $60$--$1000$ GeV. Increasing
the value of $\Lambda$ to $5$ TeV, leads to bounds that are $15$ to
$20$\% stronger, but exhibiting the same features.  Comparing Tables
\ref{tab:1} and \ref{tab:2}, we can witness that the limits on the
leptoquarks get stronger when we consider their non-universal
contributions. We can also see that the dependence on $m_{\text{top}}$
is rather weak since it is not the leading term in the $Z$ form
factors, see Eq.\ (\ref{z:ll}).

\begin{figure}
\epsfxsize=10cm
\begin{center}
\leavevmode \epsfbox{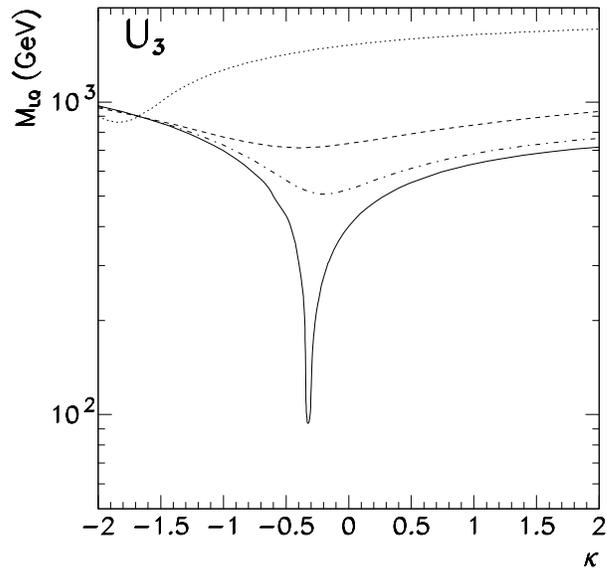}
\end{center}
\caption{Lower bounds (95 \% CL) on the mass of the vector leptoquark $U^L_3$
as a function of $\kappa=\kappa^\prime$ for 
$g_{\text{lq}}=0, \, e/s_W, \, 1, \, (4 \pi)^{1/2}$  
(solid, dot-dashed, dashed and dotted lines respectively). We considered a 
third generation 
leptoquark with $\Lambda=2$ TeV and varied all other parameters.}
\label{kappa}
\end{figure}

In order to understand the importance of the parameters $\kappa$
$\kappa^\prime$, we exhibit in Fig.\ \ref{kappa} the limits on the
mass of the leptoquark triplet $U_3$ as a function of $\kappa=
\kappa^\prime$ for several values of $g_{\text{lq}}$. As a general trend,
we can see that the limits are more restrictive as $\kappa$ and
$g_{\text{lq}}$ are increased. The position of the minimum, for a
fixed $g_{\text{lq}}$, determined by the value of $\kappa$ that
minimized the leptoquark contributions.  In fact, for low values of
$g_{\text{lq}}$ the minimum is close to $\kappa = -0.32$, which is the
value that leads to minimum oblique corrections. For large values of
$g_{\text{lq}}$, the main leptoquark effect are arises from their 
contribution to the vertex functions , and those are minimum for 
$\kappa \simeq -10/6$.

Our conclusions are that the bounds on vector leptoquarks coming from
$Z$ physics extend the limits obtained from low-energy data for second
and third generation leptoquarks. In fact, the best limit for second
generation leptoquarks come from their contribution to $(g-2)_\mu$ and
it takes the form $m_{\text{lq}} > 183~ g_{\text{lq}}$ GeV, which is
weaker than the bounds presented in Table \ref{tab:2}. Therefore, we
can conclude that our bounds exclude large regions of the parameter
space where the new collider experiments could search for these
particles, however, not all of it \cite{fut:pp,fut:ee}.  Nevertheless,
we should keep in mind that nothing substitutes the direct
observation.


\begin{center}
{\bf ACKNOWLEDGEMENTS}
\end{center}

O.J.P.E. is grateful to the Dept.\ de F\'{\i}sica Te\`orica da
Universidad de Val\`encia for its kind hospitality. This work was
supported by Conselho Nacional de Desenvolvimento Cient\'{\i}fico e
Tecnol\'ogico (CNPq-Brazil), by Funda\c{c}\~ao de Amparo \`a Pesquisa
do Estado de S\~ao Paulo (FAPESP), by DGICYT under grant PB95-1077, by
CICYT under grant AEN96-1718, and by Generalitat Valenciana and 
by EEC under the TMR contract ERBFMRX-CT96-0090.

\newpage



\protect

\protect



\protect




\protect
\begin{table}
\begin{displaymath}
\protect
\begin{array}{|l|l|}
\hline
\hline
\mbox{Quantity} & \mbox{Experimental value} \\ \hline
M_Z \mbox{[GeV]} & 91.1863 \pm 0.0020 \\
\Gamma_Z \mbox{[GeV]} & 2.4946 \pm 0.0027 \\
\sigma_{\rm had}^0  \mbox{[nb]} & 41.508 \pm 0.056\\
R_e = \frac{\Gamma({\rm had})}{\Gamma(e^+ e^-)} & 20.754 \pm 0.057 \\
R_\mu = \frac{\Gamma({\rm had})}{\Gamma(\mu^+ \mu^-)} & 20.796 \pm 0.040 \\
R_\tau = \frac{\Gamma({\rm had})}
{\Gamma(\tau^+ \tau^-)} & 20.814 \pm 0.055 \\
A_{FB}^{0e} & 0.0160 \pm 0.0024   \\
A_{FB}^{0\mu} & 0.0162 \pm 0.0013 \\
A_{FB}^{0\tau} & 0.0201 \pm 0.0018 \\
A_{\tau}^0 & 0.1401 \pm 0.0067 \\
A_e^0  & 0.1382 \pm 0.0076 \\
R_b = \frac{\Gamma(b \bar{b})}{ \Gamma({\rm had})} &0.2179 \pm 0.0012\\
R_c = \frac{\Gamma(c\bar{c}) }{\Gamma({\rm had})} & 0.1715 \pm 0.0056\\
A_{FB}^{0b} & 0.0979 \pm 0.0023  \\
A_{FB}^{0c} & 0.0733 \pm 0.0048  \\
\hline
\hline
\end{array}
\end{displaymath}
\caption{LEP data}
\label{LEPdata}
\end{table}

\widetext


%
\begin{table}
\begin{displaymath}
\begin{array}{|c|c|c|c|c|c|}
\hline
\kappa & U_1^{L(R)}  &\tilde  U_1^R & U_3^L  & V_2^{L(R)}  & \tilde V_2^L \\
\hline
0 &   70- 110 &  150- 225  &  240- 475 &  155- 300  &  130- 275 \\ 
  &   80- 120 &  175- 260  &  285- 590 &  180- 354 &  150- 320 \\ 
\hline
1 &  120- 180 &  260- 370  &  400- 750 & 250- 485  &  220- 450 \\
  &  140- 210 &  300- 450  &  475- 960 & 290- 590  &  260- 550 \\
 \hline
\end{array}
\end{displaymath}
\caption{95 \% CL lower limits on vector leptoquarks from oblique corrections. 
Upper (lower) line is for $\Lambda=2 (5)$ TeV}
\label{tab:1}
\end{table}


\begin{table}
\begin{displaymath}
\begin{array}{|c|c|c|c|c|c|c|c|c|}
\hline
\kappa & g & U_1^L  & U_1^R & \tilde  U_1^R & U_3^L & V_2^L 
& V_2^R & \tilde V^L_2 \\
\hline
  & \sqrt{4\pi} &  590- 690 &  880- 930 & 1040-1190 & 1490-1630 &
 1040-1210 & 1160-1220 & 1180-1340 \\
  & &  660- 840 &  720- 790 &  970-1050 & 1410-1560 &
  890-1060 &  950-1030 &  970-1180 \\\cline{2-9}
  & 1 &  220- 280 &  350- 370 &  430- 490 &  770- 820 &
  450- 460 &  470- 530 &  530- 610 \\
0  & &  250- 350 &  260- 310 &  360- 440 &  700- 740 &
  330- 370 &  310- 450 &  410- 470 \\\cline{2-9}
  & \frac{\displaystyle e}{\displaystyle s_W} 
  &  140- 200 &  240- 250 &  280- 310 &  450- 590 &
  200- 350 &  230- 400 &  360- 400 \\
  & &  160- 240 &  160- 210 &  200- 310 &  380- 550 &
  150- 290 &  180- 350 &  260- 360 \\
 \hline
  & \sqrt{4\pi} &  690- 810 & 1010-1060 & 1270-1320 & 1600-1730 &
 1180-1320 & 1300-1340 & 1310-1470 \\
  & &  770- 970 &  840- 920 & 1100-1180 & 1530-1670 &
 1020-1180 & 1070-1170 & 1100-1310 \\\cline{2-9}
  & 1 &  260- 340 &  410- 430 &  500- 580 &  790- 930 &
  370- 570 &  430- 650 &  630- 650 \\
1  & &  290- 410 &  300- 370 &  360- 520 &  680- 860 &
  280- 480 &  310- 570 &  460- 500 \\\cline{2-9}
  & \frac{\displaystyle e}{\displaystyle s_W} 
  &  180- 260 &  250- 290 &  270- 420 &  430- 790 &
  250- 500 &  260- 540 &  300- 520 \\
  & &  200- 300 &  170- 260 &  250- 400 &  400- 760 &
  240- 470 &  250- 510 &  230- 430 \\
 \hline
\end{array}
\end{displaymath}
\label{tab:2}
\caption{95 \% CL lower limits for second (upper lines) and 
third (lower lines) generation vector leptoquarks for $\Lambda=2$ TeV}
\end{table}


\begin{references}



\bibitem{comp} See, for instance, W.\ Buchm\"uller, Acta Phys.\
  Austriaca Suppl.\ {\bf XXVII} (1985) 517.


\bibitem{gut}See, for instance, P.\ Langacker, Phys.\ Rep.\ {\bf 72}
  (1981) 185.


\bibitem{tech}See, for instance, E.\ Farhi and L.\ Susskind, Phys.\
  Rep.\ {\bf 74} (1981) 277.


\bibitem{rizzo}See, for instance, J.\ L.\ Hewett and T.\ G.\ Rizzo,
  Phys.\ Rep.\ {\bf 183} (1989) 193.


\bibitem{sm} The LEP Electroweak Working Group, LEPEWWG/96-02.

\bibitem{nois}J.\ K.\ Mizukoshi, O.\ J.\ P.\ \'Eboli, and M.\ C.\
Gonzalez-Garcia, Nucl.\ Phys.\ {\bf B443} (1995) 20; G.\
Bhattacharyya, J.\ Ellis, and K.\ Sridhar, Phys.\ Lett.\ {\bf B336}
(1994) 100, erratum {\sl ibid.\/} {\bf B338}, 522 (1994).

\bibitem{lep} ALEPH Coll., deCamp {\sl et al.}, Phys.\ Rep.\
{\bf 216} (1992) 253; DELPHI Coll, P.\ Abreu {\sl et al.},
Phys.\ Lett.\ {\bf B316} (1993) 620; L3 Coll, B.\ Adeva {\sl
et al.}, Phys.\ Rep.\ {\bf 236} (1993) 1; OPAL Coll, G.\
Alexander {\sl et al.}, Phys.\ Lett.\ {\bf B263} (1992) 123.


\bibitem{hera}ZEUS Coll., M.\ Derrick {\sl et al.}, Phys.\
  Lett.\ {\bf B306} (1993) 173; H1 Coll, I.\ Abt {\sl et
  al.}, Nucl.\ Phys.\ {\bf B396} (1993) 3; H1 Coll, S.\ Aid
  {\sl el al.}, Phys.\ Lett.\ {\bf 369} (1996) 173.


\bibitem{pp}CDF Coll., F.\ Abe {\sl et al.}, Phys.\ Rev.\
  {\bf D48} (1993) 3939; Phys.\ Rev.\ Lett.\ {\bf 75} (1995) 1012; D0
  Coll, S.\ Abachi {\sl et al.}, Phys.\ Rev.\ Lett.\ {\bf 72}
  (1994) 965. {\sl ibid.\/} {\bf 75} (1995) 3618; K.\ Maeshima
  preprint FERMILAB-CONF-96/413-E, in the proceedings of the 28$^{th}$
  International Conference in High Energy Physics, Warsaw, 1996.


\bibitem{fut:pp} O.\ J.\ P.\ \'Eboli and J.\ E.\ Cieza Montalvo, Phys.\
   Rev.\ {\bf D50} (1994) 331; J.\ L.\ Hewett {\sl et al.}, in the
   proceedings of the Workshop on Physics at Current Accelerators and
   the Supercollider (hep-ph/9310361). J.\ Bl\"umlein, E.\ Boos, and
   A.\ Kryukov, DESY 96-174 (hep-ph/9610408); M.\ S.\ Berger and W.\
   Merrit, preprint IUHET-348 (hep-ph/9611386).




\bibitem{fut:ee}J.\ E.\ Cieza and O.\ J.\ P.\ \'Eboli, Phys.\
  Rev.\ {\bf D47} (1993) 837; J.\ Bl\"umein and R.\ R\"uckl, Phys.\
  Lett.\ {\bf B304} (1993) 337; T.\ M.\ Aliev, E.\ Iltan, and N.\ K.\
  Pak, Phys.\ Rev.\ {\bf D54} (1996) 4263; F.\ Cuypers, Int.\ J.\
  Mod.\ Phys.\ {\bf A11} (1996) 1627; J.\ Bl\"umlein, E.\ Boos, and
  A.\ Kryukov, DESY 96-219 (hep-ph/9610506).

\bibitem{shanker} O.\ Shanker, Nucl.\ Phys.\ {\bf B204} (1982) 375.

\bibitem{fcnc} W.\ Buchm\"uller and D.\ Wyler, Phys.\ Lett.\ {\bf
    B177} (1986) 377; J.\ C.\ Pati and A.\ Salam, Phys.\ Rev.\ {\bf
    D10} (1974) 275.

\bibitem{leurer}M.\ Leurer, Phys.\ Rev.\ {\bf D50} (1994) 536; 
S.\ Davidson, D.\ Bailey, and A.\ Campbell, Z.\ Phys.\ {\bf C61}
(1994) 613.  

\bibitem{apv} P.\ Langacker, Phys.\ Lett.\ {\bf B256} (1991) 277; P.\
  Langacker, M.\ Luo, and A.\ K.\ Mann, Rev.\ Mod.\ Phys.\ {\bf 64}
  (1992) 87.

\bibitem{buch} W.\ Buchm\"uller, R.\ R\"uckl, and D.\ Wyler, Phys.\ Lett.\
{\bf B191} (1987) 442.


\bibitem{hollik} See, for example, W.\ Hollik, in the Proceedings of
  the VII Swieca Summer School, editors O.\ J.\ P.\ \'Eboli and V.\
  O.\ Rivelles (World Scientific, Singapore, 1994).


\bibitem{obli} G.\ Altarelli, R.\ Barbieri, and F.\ Caravaglios,
  Nucl.\ Phys.\ {\bf B405} (1993) 3.

\bibitem{reg:dim} G.\ 't Hooft and M.\ Veltman, Nucl.\ Phys.\ {\bf B44}
  (1972) 189; C.\ G.\ Bollini and J.\ J.\ Giambiagi, Nuovo Cim.\ {\bf
    12B} (1972) 20.

\bibitem{ward} J.\ C.\ Ward, Phys.\ Rev.\ {\bf 78} (1950) 182;
Y.\ Takahashi, Nuovo Cim.\ {\bf 6} (1957) 317; J.\ C.\ Taylor, Nucl.\
Phys.\ {\bf B33} (1971) 436; A.\ A.\ Slavnov, Theor.\ Math.\ Phys.\
{\bf 10} (1972) 99.



\bibitem{zfit}D.\ Bardine {\sl et al.},  CERN-TH 6443/92 and refs.\ therein.

\bibitem{mtop} CDF Coll., F.\ Abe {\sl et al.}, 
 Int.\ J.\ Mod.\ Phys.\ {\bf A11} (1996) 2045; 
 D0 Coll., S.\ Abachi {\sl et al.}, Phys.\ Rev.\ {\bf D52}  (1995) 4877. 
\end{references}
\end{document}